\documentclass[aps,prb,reprint,superscriptaddress,floatfix]{revtex4-2}

\usepackage{amsmath,amssymb,graphicx,bm}
\usepackage{hyperref}
\usepackage{placeins}

\begin{document}

\title{Electrical Spectroscopy of Intervalley Relaxation in WSe$_2$ Transistors}

\author{Katsunori Wakabayashi}
\email[Email Address: ]{WAKABAYASHI.Katsunori@nims.go.jp}
\affiliation{Research Center for Materials Nanoarchitectonics (MANA), 
National Institute for Materials Science (NIMS),
Tsukuba 305-0044, Japan}
\affiliation{Kwansei Gakuin University, Gakuen-Uegahara 1, Sanda 669-1330, Japan}

\date{\today}

\begin{abstract}
We show that the transconductance of multilayer WSe$_2$ field-effect
transistors serves as a direct electrical spectrometer of the
intervalley relaxation time $\tau_{\rm iv}$, previously accessible only
by ultrafast optical techniques.
Extending an equilibrium valley-thermodynamics framework with a single
relaxation equation for the $\Gamma$-valley carrier fraction $f_\Gamma(t)$,
we predict three signatures:
(i)~a Lorentzian transconductance
$g_m(\omega)=g_{m,0}+g_{m,v}^0/(1+i\omega\tau_{\rm iv})$,
whose imaginary part peaks at $\omega_c=\tau_{\rm iv}^{-1}$ with
opposite signs for bilayer and trilayer;
(ii)~a two-stage current transient after a gate step, exhibiting
bilayer overshoot or trilayer undershoot; and
(iii)~sweep-rate-proportional hysteresis whose gate-voltage profile
and the layer-dependent transconductance sign reversal distinguish
valley from trap-induced dynamics.
All three signatures provide quantitative electrical access to
$\tau_{\rm iv}$ with standard radio-frequency (rf) and
direct-current (dc) instrumentation.
\end{abstract}

\maketitle

\section{Introduction}

A transistor channel hosting multiple valleys with different effective
masses carries a slow internal degree of freedom: the intervalley
carrier distribution.
When the intervalley equilibration time $\tau_{\rm iv}$ is not
negligibly short compared with the gate-modulation period, this
distribution cannot follow the gate adiabatically, imprinting on
all time-dependent transport observables.
The consequences form a recognizable set of dynamic signatures ---
frequency-dependent transconductance, two-stage current transients,
and sweep-rate-proportional hysteresis --- whose gate-voltage profiles
and the layer-dependent transconductance sign reversal distinguish
valley dynamics from charge-trapping effects.
Accessing $\tau_{\rm iv}$ through standard rf and dc instrumentation
would provide a direct spectroscopic route to an internal scattering
channel previously confined to ultrafast optical laboratories.

Multilayer WSe$_2$ provides an ideal, layer-programmable platform for
the study of these effects~\cite{Wang2012electronics,Manzeli2017TMDCs,Mak2018_valleyreview,Schaibley2016_valleytronics,Radisavljevic2011MoS2,Fang2012WSe2FET,Movva2015WSe2,Kim2024_WSe2contact}.
The valence-band maximum shifts from the $K$ point in the monolayer
toward the $\Gamma$ point with increasing layer number, driven by
interlayer coupling and spin-orbit
interaction~\cite{Zhao2013_WSe2layers,Zhang2014_WSe2ARPES,Wickramaratne2014_indirect,Zhu2011spinorbit,Liu2013tightbinding},
placing $\Delta_{K\Gamma}\approx k_BT$ near room temperature in the
bilayer.
This competition enables gate-tunable redistribution between the
$K$ and $\Gamma$ valleys, producing a valley-dependent transconductance
contribution $g_{m,v}=-\Delta\mu\,(W/L)V_{DS}qp\chi_v$ parameterized by
the valley susceptibility $\chi_v\equiv\partial f_\Gamma/\partial
V_{GS}$, completing the decomposition $g_m =
g_{m,0}+g_{m,v}$~\cite{PaperI,PaperII,Xiao2012_valley,Mak2012valley,Zeng2012valley,Mak2014valleyHall}.
Such treatments assumed quasi-static valley equilibrium.
The relaxation time $\tau_{\rm iv}$ in WSe$_2$ is set by phonon-mediated
scattering at room temperature (sub-ps to a few
ps)~\cite{Lee2021_phonon,Bae2022_intervalley,Kioseoglou2012intervalley,Schmidt2016intervalley,DalConte2015ultrafast,Oh2023_WSe2multilayer,Dogadov2026_intervalley}
but can be substantially longer at reduced temperatures or under
gate-induced non-equilibrium
conditions~\cite{Wang2014_valleyrelax,Bertoni2016_dynamics,Glazov2014spinvalley},
potentially placing $\omega_c=\tau_{\rm iv}^{-1}$ in the accessible
MHz--GHz window.
To date, $\tau_{\rm iv}$ in WSe$_2$ has been probed almost exclusively
by ultrafast optical spectroscopy~\cite{Madeo2020darkexciton,Wallauer2021_trARPES}, which
requires specialized laser and optical cryostat infrastructure beyond
conventional electrical transport setups.

\begin{figure*}[t]
  \includegraphics[width=\textwidth]{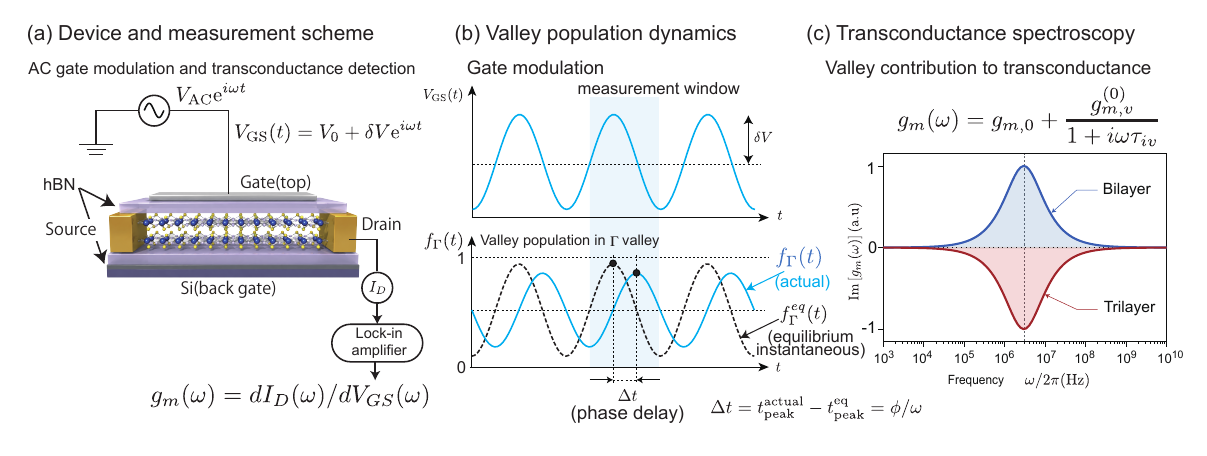}
  \caption{%
    Overview of transconductance spectroscopy of intervalley relaxation.
    (a)~Schematic of a multilayer WSe$_2$ transistor in dual-gate geometry
    with hexagonal boron nitride (hBN) dielectric~\cite{Geim2013vdW,Dean2010hBN}.
    An alternating-current (ac) gate voltage $V_{GS}(t)=V_0+\delta V e^{i\omega t}$
    drives the device; the drain current $I_D(\omega)$ is detected by a
    lock-in amplifier to extract
    $g_m(\omega)=dI_D(\omega)/dV_{GS}(\omega)$.
    (b)~Valley population $f_\Gamma(t)$ in the $\Gamma$ valley (solid)
    compared to the instantaneous equilibrium value (dashed), illustrating
    the phase delay $\Delta t = \phi/\omega$ arising from the finite
    intervalley relaxation time $\tau_{\rm iv}$.
    (c)~Predicted imaginary part of the valley transconductance
    $\mathrm{Im}[g_m(\omega)]$ for bilayer (blue) and trilayer (red)
    WSe$_2$.
    Opposite-sign Lorentzian peaks centered at $\omega_c=\tau_{\rm iv}^{-1}$
    provide a direct spectroscopic readout of $\tau_{\rm iv}$.
  }
  \label{fig:overview}
\end{figure*}

In this paper, we extend the equilibrium valley framework to the
non-equilibrium regime through a single relaxation equation for
$f_\Gamma(t)$, characterized by the intervalley relaxation time
$\tau_{\rm iv}$.
This minimal extension predicts three measurable consequences of
delayed intervalley redistribution:
(i) a Lorentzian frequency dependence of $g_m(\omega)$ with
characteristic frequency $\omega_c = \tau_{\rm iv}^{-1}$;
(ii) a two-stage current transient after a gate step exhibiting
bilayer overshoot or trilayer undershoot, determined by the sign of
$g_{m,v}^0$;
and (iii) sweep-rate-proportional hysteresis whose gate-voltage profile
follows $qp\chi_v(V_{GS})$ and reverses sign between the two layer
numbers.
In each case, the transconductance sign reversal with layer number and
the gate-voltage dependence allow valley-origin dynamics to be
distinguished from charge-trapping effects, and together they provide
a quantitative electrical route to $\tau_{\rm iv}$.
Figure~\ref{fig:overview}(a) shows the device geometry and
measurement configuration, Fig.~\ref{fig:overview}(b) illustrates
the resulting valley-population lag, and Fig.~\ref{fig:overview}(c)
displays the predicted spectroscopic signature in $\mathrm{Im}[g_m(\omega)]$.

\begin{figure}[!t]
  \includegraphics[width=0.95\columnwidth]{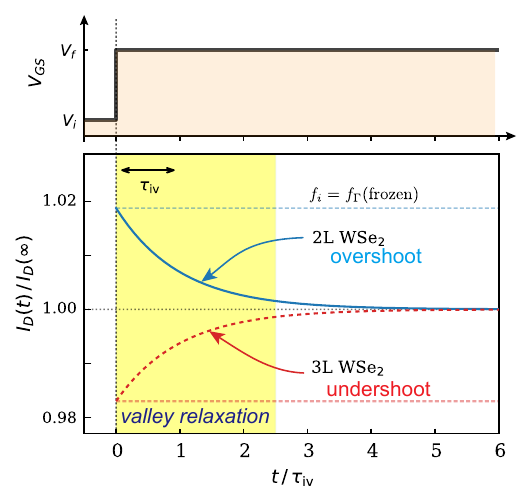}
  \caption{%
    Two-stage current response to a gate step $\Delta V_{GS}=0.2\,\mathrm{V}$
    (upper panel) for bilayer (blue) and trilayer (red) WSe$_2$ (lower panel,
    thick lines).
    Thin dashed lines ($f_\Gamma=f_i$, frozen) show the valley-frozen
    reference: charge responds instantaneously ($\tau_{RC}\ll\tau_{\rm iv}$)
    while $f_\Gamma$ remains at its pre-step value $f_i$.
    Valley relaxation then drives the current toward $I_D(\infty)$:
    overshoot for the bilayer and undershoot for the trilayer, with a
    characteristic timescale $\tau_{\rm iv}$.
    The opposite-sign departures reflect $\mathrm{sgn}(g_{m,v}^0)$
    reversing between layers.
  }
  \label{fig:concept}
\end{figure}
\section{Dynamic Valley Model}
\label{sec:model}

The equilibrium valley-thermodynamics framework~\cite{PaperII} yields
a transconductance decomposition $g_m^{(0)} = g_{m,0}+g_{m,v}^0$,
where $g_{m,0}$ is the conventional charge-accumulation term and
\begin{equation}
  g_{m,v}^0 = -\frac{W}{L}V_{DS}\,\Delta\mu\,qp\,\chi_v,
  \label{eq:gm_static}
\end{equation}
with $\Delta\mu=\mu_K-\mu_\Gamma>0$, $q$ the elementary charge,
$p$ the total hole density, and
$\chi_v\equiv\partial f_\Gamma/\partial V_{GS}$ the valley susceptibility.
The sign of $g_{m,v}^0$ follows that of $\Delta_{K\Gamma}$: negative
for the bilayer, positive for the trilayer.

Here $f_\Gamma^{\rm eq}(V_{GS})$ is the equilibrium fraction of holes
occupying the $\Gamma$ valley at gate voltage $V_{GS}$:
\begin{equation}
  f_\Gamma^{\rm eq}(V_{GS}) =
    \frac{p_\Gamma^{\rm eq}(V_{GS})}{p_K^{\rm eq}(V_{GS})+p_\Gamma^{\rm eq}(V_{GS})},
  \label{eq:feq}
\end{equation}
where $p_\nu^{\rm eq}$ is the equilibrium two-dimensional hole density
in valley $\nu\in\{K,\Gamma\}$, determined from Fermi--Dirac statistics
with valley density of states $D_\nu\propto m_\nu^*$ and the
gate-controlled Fermi energy~\cite{PaperII}.
The susceptibility $\chi_v$ peaks near the $K$--$\Gamma$ band-crossing
condition and vanishes in both the subthreshold ($p\to 0$) and heavily
accumulated limits, confining the valley contribution
$g_{m,v}^0\neq 0$ to the above-threshold partially occupied regime.

\textit{Layer-dependent sign reversal.}---
The sign of $\chi_v$---and hence of $g_{m,v}^0$---is determined by
which valley constitutes the valence-band maximum, a property that
reverses sign with layer number in
WSe$_2$~\cite{Zhao2013_WSe2layers,Wickramaratne2014_indirect}.
In the bilayer ($\Delta_{K\Gamma}=+26\,\mathrm{meV}$), the $K$ valley
lies at higher hole energy: holes at low carrier density reside
predominantly in $K$, and a positive gate step drives them
progressively into $\Gamma$, giving $\chi_v>0$ and hence
$g_{m,v}^0<0$.
In the trilayer ($\Delta_{K\Gamma}=-49\,\mathrm{meV}$), the ordering
reverses: holes first fill $\Gamma$, and the $\Gamma$ fraction
decreases as $K$ also becomes occupied with increasing gate voltage,
giving $\chi_v<0$ and $g_{m,v}^0>0$.
This layer-dependent transconductance sign reversal is absent in
single-valley materials or in systems where $\Delta_{K\Gamma}$ has a
fixed sign, and is specific to the layer-dependent band structure
of multilayer WSe$_2$.

We extend this to the nonequilibrium regime with a single relaxation
equation for the $\Gamma$-valley fraction $f_\Gamma(t)=p_\Gamma(t)/p(t)$:
\begin{equation}
  \frac{df_\Gamma}{dt}
  = \frac{f_\Gamma^{\rm eq}[V_{GS}(t)] - f_\Gamma(t)}{\tau_{\rm iv}},
  \label{eq:relax}
\end{equation}
the transistor analogue of Debye relaxation in
dielectrics~\cite{Cole1941_dielectric}.
Extensions to distributed $\tau_{\rm iv}$ (Cole--Cole, stretched
exponential) follow directly. The single-$\tau_{\rm iv}$ case is
the falsifiable baseline.

\textit{Separation of timescales.}---
Equation~(\ref{eq:relax}) is valid in the regime
$\tau_{RC}\ll\tau_{\rm iv}$, where the charge relaxation time
$\tau_{RC}=R_{ch}C_{ox}$ is much shorter than the intervalley
relaxation time.
Under this assumption, the total hole density $p(t)=p[V_{GS}(t)]$
tracks $V_{GS}$ instantaneously, while $f_\Gamma(t)$ lags.
The source and drain contacts are assumed to act as thermal reservoirs
that define the carrier injection distribution at the channel
boundaries, while transport within the channel is treated in the
low-bias diffusive limit~\cite{Silva2026_memtransistor,Jimenez2012_driftdiff}.
The channel hole density $p(V_{GS})=C_{\rm ox}(V_{GS}-V_{\rm th})/q$,
with $V_{\rm th}$ the threshold voltage, responds electrostatically to the gate.
The intervalley fraction $f_\Gamma(t)$ modulates the effective
channel mobility
\begin{equation}
  \mu_{\rm eff} = \mu_K(1-f_\Gamma)+\mu_\Gamma f_\Gamma
                = \mu_K-\Delta\mu\,f_\Gamma,
  \label{eq:mueff}
\end{equation}
but leaves $p$ unchanged by intervalley redistribution.
This distinguishes the present mechanism from charge-trapping
memtransistors, in which the trapped charge renormalizes the channel
band edge and gate efficiency~\cite{Silva2026_memtransistor}. In
the present model, no additional interface capacitance or band-edge
shift is introduced.
The dynamic drain current is then
\begin{equation}
  I_D(t)=\frac{W}{L}\,\mu_{\rm eff}[f_\Gamma(t)]\,qp[V_{GS}(t)]\,V_{DS}.
  \label{eq:ID_dynamic}
\end{equation}
Figure~\ref{fig:concept} illustrates the physical picture:
a gate-voltage step ($V_{GS}:V_1\to V_2$) causes the charge to jump
instantaneously while $f_\Gamma$ exponentially relaxes toward
the new equilibrium, producing a two-stage current response.

Table~\ref{tab:params} lists the numerical parameters used throughout
this work.
The effective masses and $K$--$\Gamma$ splittings are taken from
first-principles calculations~\cite{Wickramaratne2014_indirect,Liu2013tightbinding},
and the mobilities are representative experimental
values~\cite{Fang2012WSe2FET,Movva2015WSe2}.
The intervalley relaxation time $\tau_{\rm iv}$ enters only as a free
parameter; no specific numerical value is assumed in any figure.

\begin{table*}[!t]
  \caption{Parameters used in the two-valley model.}
  \label{tab:params}
  \begin{ruledtabular}
  \begin{tabular}{lll}
    Quantity & Value & Description \\
    \hline
    $m_K^*$                & $0.40\,m_0$                      & $K$-valley effective mass \\
    $m_\Gamma^*$           & $1.00\,m_0$                      & $\Gamma$-valley effective mass \\
    $\Delta_{K\Gamma}$(2L) & $+26\,\mathrm{meV}$              & $K$--$\Gamma$ valley splitting, bilayer \\
    $\Delta_{K\Gamma}$(3L) & $-49\,\mathrm{meV}$              & $K$--$\Gamma$ valley splitting, trilayer \\
    $\mu_K$                & $100\,\mathrm{cm^2/(V{\cdot}s)}$ & $K$-valley hole mobility \\
    $\mu_\Gamma$           & $30\,\mathrm{cm^2/(V{\cdot}s)}$  & $\Gamma$-valley hole mobility \\
    $C_{\rm ox}$           & $49\,\mathrm{mF/m^2}$            & Gate oxide capacitance per unit area\footnote{Equivalent oxide thickness $0.7\,\mathrm{nm}$.} \\
    $W = L$                & $1\,\mu\mathrm{m}$               & Channel width and length \\
    $V_{DS}$               & $50\,\mathrm{mV}$                & Drain-source bias voltage \\
    $T$                    & $300\,\mathrm{K}$                & Temperature \\
  \end{tabular}
  \end{ruledtabular}
\end{table*}

\begin{figure}[!t]
  \includegraphics[width=0.95\columnwidth]{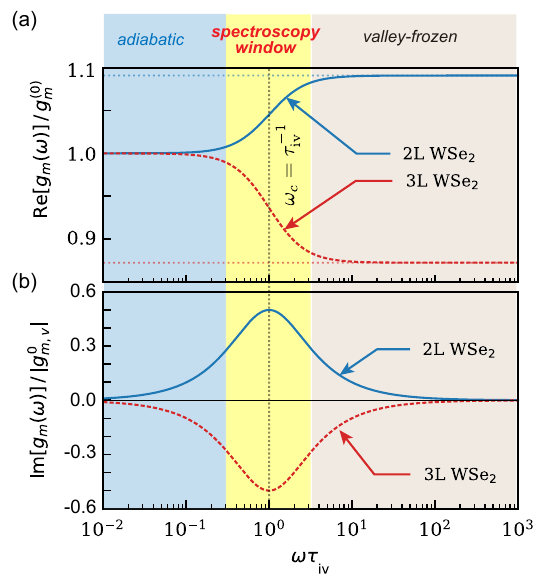}
  \caption{%
    Frequency-dependent transconductance for bilayer (solid, blue) and
    trilayer (dashed, red) WSe$_2$ at $T=300\,\mathrm{K}$,
    evaluated at the gate voltage where $|g_{m,v}^0|$ is maximum.
    Shaded regions mark three dynamical regimes: adiabatic
    ($\omega\tau_{\rm iv}\ll 1$), spectroscopy window
    ($\omega\tau_{\rm iv}\sim 1$, yellow), and valley-frozen
    ($\omega\tau_{\rm iv}\gg 1$).
    (a)~Real part $\mathrm{Re}[g_m(\omega)]/g_m^{(0)}$ normalized to
    the dc value; dotted lines are the high-frequency
    limits $g_{m,0}/g_m^{(0)}$.
    (b)~Imaginary part $\mathrm{Im}[g_m(\omega)]/|g_{m,v}^0|$;
    the extremum at $\omega_c=\tau_{\rm iv}^{-1}$ (vertical dotted)
    has opposite sign for bilayer ($g_{m,v}^0<0$) and trilayer
    ($g_{m,v}^0>0$).
  }
  \label{fig:freq_response}
\end{figure}
\section{Frequency-Dependent Transconductance}
\label{sec:freq}
For a small ac modulation $V_{GS}(t)=V_0+\delta V e^{i\omega t}$,
linearizing Eq.~(\ref{eq:relax}) gives the
\emph{dynamic valley susceptibility}
\begin{equation}
  \chi_v(\omega;V_0) \equiv \frac{\delta f_\Gamma}{\delta V}
  = \frac{\chi_v(V_0)}{1+i\omega\tau_{\rm iv}},
  \label{eq:chiv_dynamic}
\end{equation}
a Lorentzian whose pole is at $\omega_c=\tau_{\rm iv}^{-1}$.
Substituting into the linearized current of Eq.~(\ref{eq:ID_dynamic}):
\begin{equation}
  g_m(\omega) = g_{m,0}(V_0)
    + \frac{g_{m,v}^0(V_0)}{1+i\omega\tau_{\rm iv}}.
  \label{eq:gm_omega}
\end{equation}
The charge-accumulation term $g_{m,0}$ is frequency-independent
(instantaneous charge relaxation); all dispersion is carried by the
valley term.
Equation~(\ref{eq:gm_omega}) is the central result of this paper.

Separating real and imaginary parts:
\begin{align}
  \mathrm{Re}[g_m(\omega)]
  &= g_{m,0} + \frac{g_{m,v}^0}{1+(\omega\tau_{\rm iv})^2},
  \label{eq:Re_gm}\\[3pt]
  \mathrm{Im}[g_m(\omega)]
  &= -\frac{g_{m,v}^0\,\omega\tau_{\rm iv}}{1+(\omega\tau_{\rm iv})^2}.
  \label{eq:Im_gm}
\end{align}
Three limiting behaviors follow immediately:
(i)~at $\omega\tau_{\rm iv}\ll 1$,
$\mathrm{Re}[g_m]=g_m^{(0)}$ recovers the equilibrium anomaly of
Ref.~\onlinecite{PaperII};
(ii)~at $\omega\tau_{\rm iv}\gg 1$,
$\mathrm{Re}[g_m]\to g_{m,0}$, the anomaly is completely suppressed;
(iii)~at $\omega=\omega_c$,
$|\mathrm{Im}[g_m]|$ reaches its maximum of $|g_{m,v}^0|/2$,
providing a sharp spectroscopic feature that directly identifies
$\tau_{\rm iv}$.
Because $g_{m,v}^0<0$ for the bilayer and $g_{m,v}^0>0$ for the
trilayer, the imaginary part peaks with \emph{opposite signs} for the
two systems, providing an additional layer-number signature.

These Kramers--Kronig-related dispersive and absorptive components are
shown in Figs.~\ref{fig:freq_response}(a) and \ref{fig:freq_response}(b),
respectively.
The high-frequency limit $g_{m,0}/g_m^{(0)}$ exceeds unity for the
bilayer (the valley suppression disappears) and falls below unity for
the trilayer (the valley enhancement disappears), directly reflecting
the sign of $g_{m,v}^0$ [Fig.~\ref{fig:freq_response}(a)].
A frequency sweep at fixed $V_{GS}=V_{GS}^*$ (the peak-$|g_{m,v}^0|$
voltage) yields $\omega_c$ from either the $-3\,\mathrm{dB}$ rolloff
of $\mathrm{Re}[g_m]-g_{m,0}$ or the peak of $|\mathrm{Im}[g_m]|$,
with both criteria giving the same $\tau_{\rm iv}$
[Fig.~\ref{fig:freq_response}(b)].

\begin{figure*}[t]
  \includegraphics[width=\textwidth]{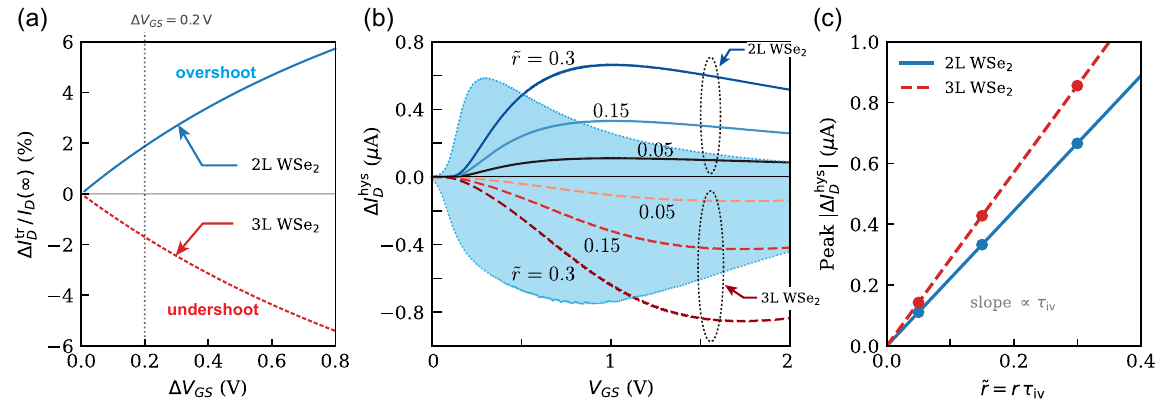}
  \caption{%
    Low-frequency fingerprints of delayed intervalley relaxation.
    (a)~Signed transient amplitude $\Delta I_D^{\rm tr}/I_D(\infty)$
    vs gate-step size $\Delta V_{GS}$ at $T=300\,\mathrm{K}$,
    evaluated at the gate voltage of maximum $|\chi_v|$.
    Bilayer (blue) overshoots and trilayer (red) undershoots, encoding
    $\mathrm{sgn}(g_{m,v}^0)$; amplitudes increase linearly before saturating.
    The dotted line marks the $\Delta V_{GS}=0.2\,\mathrm{V}$ step of
    Fig.~\ref{fig:concept}.
    (b)~Hysteresis current $\Delta I_D^{\rm hys}(V_{GS})$
    [Eq.~(\ref{eq:hys_current})] for bilayer (solid, blue family) and
    trilayer (dashed, red family) at normalized sweep rates
    $\tilde{r}=r\tau_{\rm iv}=0.05$, 0.15, 0.30 (labeled).
    Opposite signs reflect $\mathrm{sgn}(\chi_v)=\mathrm{sgn}(\Delta_{K\Gamma})$;
    cyan dotted envelopes trace $\chi_v(V_{GS})$, confirming the
    profile correspondence.
    (c)~Peak $|\Delta I_D^{\rm hys}|$ vs $\tilde{r}$. Linear slopes
    confirm Eq.~(\ref{eq:hys_current}) and give $\tau_{\rm iv}$ directly
    via Eq.~(\ref{eq:tau_extract}).
  }
  \label{fig:fingerprints}
\end{figure*}
\section{Transient Response and Hysteresis}
\label{sec:transient}

\textit{Step response.}---
For a step $V_{GS}:V_i\to V_f$ at $t=0$, the charge density
jumps to $p(V_f)$ instantaneously while $f_\Gamma$ satisfies
Eq.~(\ref{eq:relax}) with initial condition
$f_\Gamma(0^-)=f_\Gamma^{\rm eq}(V_i)\equiv f_i$.
The solution is $f_\Gamma(t)=f_f+(f_i-f_f)e^{-t/\tau_{\rm iv}}$
with $f_f\equiv f_\Gamma^{\rm eq}(V_f)$, giving
\begin{equation}
  I_D(t) = I_D(\infty) + \Delta I_D^{\rm tr}\,e^{-t/\tau_{\rm iv}},
  \label{eq:ID_step}
\end{equation}
where $I_D(\infty)$ is the equilibrium current at $V_f$ and
the transient amplitude is
\begin{equation}
  \frac{\Delta I_D^{\rm tr}}{I_D(\infty)}
  = \frac{\Delta\mu}{\mu_{\rm eff}(V_f)}\,(f_f-f_i).
  \label{eq:transient_norm}
\end{equation}
The sign of $\Delta I_D^{\rm tr}$ is governed by the layer number:
for bilayer WSe$_2$, a positive gate step drives holes from the light $K$
toward the heavy $\Gamma$ valley ($f_f>f_i$), so $\Delta I_D^{\rm tr}>0$
and the current \emph{overshoots} before decaying toward equilibrium.
For trilayer, the reverse transfer gives $f_f<f_i$,
$\Delta I_D^{\rm tr}<0$, and the current \emph{undershoots} before
recovering toward equilibrium.
This layer-dependent directionality is a direct consequence of the
sign rule $\mathrm{sgn}(g_{m,v}^0)=-\mathrm{sgn}(\Delta_{K\Gamma})$
and constitutes an unambiguous experimental fingerprint
[Fig.~\ref{fig:fingerprints}(a)].
At the optimal gate voltage near the $\chi_v$ peak, the transient
amplitude can reach several percent of $I_D(\infty)$, growing
linearly with $\Delta V_{GS}$ for small steps.

\textit{Hysteresis under finite-rate gate sweeps.}---
For a linearly varying gate voltage at rate $r=dV_{GS}/dt$,
the relaxation equation~(\ref{eq:relax}) gives, to first order in
$r\tau_{\rm iv}$,
\begin{equation}
  f_\Gamma(t) \approx f_\Gamma^{\rm eq}[V_{GS}(t)]
    - r\tau_{\rm iv}\,\chi_v[V_{GS}(t)].
  \label{eq:fGamma_sweep}
\end{equation}
The valley fraction lags the equilibrium value by
$\delta f_\Gamma = -r\tau_{\rm iv}\chi_v$.
Substituting into Eq.~(\ref{eq:ID_dynamic}), the first-order current
deviation is
\begin{equation}
  \delta I_D(V_{GS}) = \frac{W}{L}\,qp\,V_{DS}\,\Delta\mu
    \,\tau_{\rm iv}\,r\,\chi_v(V_{GS}).
  \label{eq:delta_ID}
\end{equation}
This approximation is valid for $\tilde{r}\equiv r\tau_{\rm iv}\ll 1$
on the gate-voltage scale over which $\chi_v$ varies significantly.
The hysteresis current---the difference between forward ($r>0$) and
reverse ($r<0$) sweeps at the same $V_{GS}$---is
\begin{equation}
  \Delta I_D^{\rm hys}(V_{GS})
  = 2\,\frac{W}{L}\,qp\,V_{DS}\,\Delta\mu\,\tau_{\rm iv}\,|r|\,\chi_v(V_{GS}).
  \label{eq:hys_current}
\end{equation}

Equation~(\ref{eq:hys_current}) encodes several signatures that
distinguish valley-origin hysteresis from trap-induced effects:

\textbf{(i) Profile.}
$\Delta I_D^{\rm hys}\propto qp\,\chi_v(V_{GS})$
activates only above threshold (where $\chi_v\neq 0$) and tracks
the gate-voltage shape of the valley susceptibility.
Trap hysteresis follows the trap energy distribution and need not
correlate with $\chi_v(V_{GS})$.

\textbf{(ii) Layer-dependent transconductance sign reversal.}
$\Delta I_D^{\rm hys}$ is positive for the bilayer ($\chi_v>0$)
and negative for the trilayer ($\chi_v<0$).
No conventional extrinsic mechanism produces this layer-dependent
transconductance sign reversal without structural differences between
the two devices.

\textbf{(iii) Linear sweep-rate dependence.}
Valley hysteresis grows as $|r|$, enabling extraction of
\begin{equation}
  \tau_{\rm iv} = \frac{|\Delta I_D^{\rm hys}|}{2|r|\,(W/L)qp\,V_{DS}\Delta\mu\,|\chi_v|}
  \label{eq:tau_extract}
\end{equation}
from the slope of peak $|\Delta I_D^{\rm hys}|$ vs $|r|$,
using $\chi_v$ obtained independently from a static $g_m$--$V_{GS}$
measurement~\cite{PaperII} or from a first-principles
estimate~\cite{Wickramaratne2014_indirect,Liu2013tightbinding}.
Trap hysteresis typically follows a logarithmic or power-law dependence
on sweep rate~\cite{Ghatak2015_trapping,Illarionov2020_Ditreview,Karl2025_hysteresis}.

\textbf{(iv) Subthreshold swing invariance.}
$\Delta I_D^{\rm hys}=0$ in the subthreshold regime where
$\chi_v=0$~\cite{PaperII}.
Trap-induced hysteresis is often largest just below threshold where
trap filling is most active.

Experimentally, the finite-rate sweep protocol corresponds to a
periodic triangular gate-voltage waveform, as commonly employed in
cyclic gate-sweep measurements~\cite{Karl2025_hysteresis}.
After initial transients of order $\tau_{\rm iv}$ following the onset
of the waveform, $f_\Gamma(t)$ settles into a reproducible steady-state
cycle synchronized to the triangular waveform.
In this steady-state limit, each upward and downward branch is locally
described by Eq.~(\ref{eq:fGamma_sweep}), and
Eq.~(\ref{eq:hys_current}) directly characterizes the steady-state
hysteresis loop width.
This steady-state interpretation renders $\tau_{\rm iv}$ extractable
from the slope of peak $|\Delta I_D^{\rm hys}|$ versus $|r|$ without
any dependence on initial conditions.

Figures~\ref{fig:fingerprints}(b,c) show the predicted steady-state
hysteresis profiles $\Delta I_D^{\rm hys}(V_{GS})$ for bilayer and
trilayer WSe$_2$ at three normalized sweep rates
$\tilde{r}=r\tau_{\rm iv}$, together with the linear scaling of the
peak amplitude.
The grey envelope traces $qp\,\chi_v(V_{GS})$,
confirming the predicted correspondence.

\section{Temperature Dependence and Spectroscopy Window}
\label{sec:temperature}

\begin{figure}[!t]
  \includegraphics[width=0.95\columnwidth]{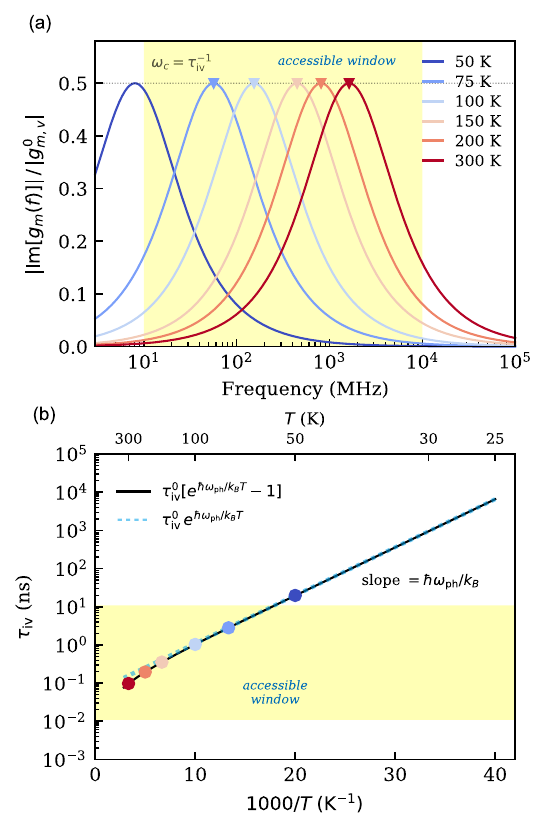}
  \caption{%
    Transconductance spectroscopy window and Arrhenius analysis.
    (a)~$|\mathrm{Im}[g_m(f)]|/|g_{m,v}^0|$ vs frequency at
    $T=50$--$300\,\mathrm{K}$ (blue to red), computed using
    Eq.~(\ref{eq:tau_T}) with $\hbar\omega_{\rm ph}=25\,\mathrm{meV}$
    and $\tau_{\rm iv}^0=60\,\mathrm{ps}$.
    Triangles mark the characteristic frequency $\omega_c/2\pi$;
    the shaded band is the accessible window
    $10\,\mathrm{MHz}$--$10\,\mathrm{GHz}$.
    (b)~Arrhenius plot of $\tau_{\rm iv}$ vs $1000/T$; the slope of the
    low-temperature regime yields $\hbar\omega_{\rm ph}/k_B$ from a
    purely electrical measurement [Eq.~(\ref{eq:Arrhenius})].
    Colored dots correspond to the temperatures in~(a).
  }
  \label{fig:spectroscopy}
\end{figure}

All three signatures are amplified at reduced temperature, and the
spectroscopy window shifts into the range of standard laboratory
instruments.
The valley susceptibility $\chi_v$ grows toward the intrinsic bound
$(4k_BT)^{-1}$ as $T$ decreases~\cite{PaperII}, while phonon-mediated
intervalley scattering slows as
\begin{equation}
  \tau_{\rm iv}(T) = \tau_{\rm iv}^0
  \bigl[e^{\hbar\omega_{\rm ph}/k_BT}-1\bigr]
  \label{eq:tau_T}
\end{equation}
for a zone-edge phonon with $\hbar\omega_{\rm ph}\approx 25\,\mathrm{meV}$
in WSe$_2$~\cite{Lee2021_phonon,Jin2014transport}.
This Bose--Einstein form arises because phonon absorption saturates at
low $T$, leaving stimulated emission as the rate-limiting
step~\cite{Kioseoglou2012intervalley,Schmidt2016intervalley}.
With $\tau_{\rm iv}^0\approx 60\,\mathrm{ps}$, this model gives
$\tau_{\rm iv}(100\,\mathrm{K})\approx 1\,\mathrm{ns}$,
placing the spectroscopic peak at
$\omega_c/2\pi\approx 160\,\mathrm{MHz}$---within the bandwidth of
commercial lock-in amplifiers and vector network analyzers.
Below $50\,\mathrm{K}$, $\omega_c/2\pi$ drops below $30\,\mathrm{MHz}$,
and remains accessible with standard instrumentation.
For comparison, ultrafast optical spectroscopy reports
$\tau_{\rm iv}\sim 0.1$--$1\,\mathrm{ps}$ at room temperature in
WSe$_2$~\cite{Lee2021_phonon,Bae2022_intervalley,Oh2023_WSe2multilayer},
placing $\omega_c/2\pi$ in the $0.1$--$10\,\mathrm{THz}$ range,
far above standard electrical bandwidths.
The present electrical method becomes viable precisely because
phonon-mediated intervalley scattering slows substantially at reduced
temperature, shifting $\omega_c$ into the MHz--GHz window demonstrated
in Fig.~\ref{fig:spectroscopy}(a).
Figure~\ref{fig:spectroscopy}(a) maps this accessible window,
with the shaded band marking the range $10\,\mathrm{MHz}$--$10\,\mathrm{GHz}$.

An Arrhenius fit of $\ln\tau_{\rm iv}$ vs $1/T$ within this window
yields the phonon coupling energy directly:
\begin{equation}
  \frac{d\ln\tau_{\rm iv}}{d(1/T)}
  \xrightarrow{k_BT\ll\hbar\omega_{\rm ph}}
  \frac{\hbar\omega_{\rm ph}}{k_B},
  \label{eq:Arrhenius}
\end{equation}
providing a purely electrical determination of the zone-edge phonon
coupling energy [Fig.~\ref{fig:spectroscopy}(b)].
The parameters $\tau_{\rm iv}^0$ and $\hbar\omega_{\rm ph}$ can be
extracted from a two-parameter fit to the measured $\omega_c(T)$
curve without any optical access.

\section{Discussion}
\label{sec:discussion}
These three phenomena provide complementary routes to determine
$\tau_{\rm iv}$ experimentally.
In ac measurements, the frequency sweep at fixed $V_{GS}^*$ gives
$\tau_{\rm iv}$ from the peak of $|\mathrm{Im}[g_m(\omega)]|$ or the
$-3\,\mathrm{dB}$ rolloff of $\mathrm{Re}[g_m(\omega)]-g_{m,0}$,
both pointing to the same $\omega_c$.
In transient measurements, the single-exponential decay of
$I_D(t)-I_D(\infty)$ directly yields $\tau_{\rm iv}$.
In gate-sweep hysteresis measurements, the linear-in-$|r|$ slope via
Eq.~(\ref{eq:tau_extract}) requires only a static $g_m$--$V_{GS}$ curve
and a rate-dependent cyclic sweep, with no rf electronics.

The quasi-equilibrium limit of Ref.~\onlinecite{PaperII} corresponds to
$\tau_{\rm iv}\ll\tau_{\rm tr}$, where $\tau_{\rm tr}=L^2/(\mu_K V_{DS})$
is the carrier transit time in the diffusive transport regime.
For $W=L=1\,\mu\mathrm{m}$, $V_{DS}=50\,\mathrm{mV}$, and
$\mu_K=100\,\mathrm{cm^2\,V^{-1}\,s^{-1}}$~\cite{Jin2014transport,Kaasbjerg2012phonon},
the transit frequency is
$\tau_{\rm tr}^{-1}\approx 80\,\mathrm{MHz}$.
If $\tau_{\rm iv}\sim\tau_{\rm tr}$, the crossover in
$\mathrm{Re}[g_m(\omega)]$ falls in the $10$--$100\,\mathrm{MHz}$ range,
accessible with commercial impedance analyzers.
Shorter-channel devices reduce $\tau_{\rm tr}$.
In the ballistic limit $\tau_{\rm tr}$ is replaced by $L/v_{\rm inj}$
(with $v_{\rm inj}$ the thermal injection velocity), further shifting
$\omega_c$ to higher frequencies and providing an additional experimental
degree of freedom through channel-length scaling of the anomaly.
Since $\tau_{\rm iv}$ is a bulk material constant set by electron-phonon
coupling, it is independent of channel geometry.
Accordingly, all three normalized observables
($\Delta I_D^{\rm tr}/I_D(\infty)$, $|\mathrm{Im}[g_m]|/|g_{m,v}^0|$,
and $\Delta I_D^{\rm hys}/I_D$) are geometry-invariant.
Reducing $L$ at fixed $W/L$ further relaxes the
$\tau_{RC}\ll\tau_{\rm iv}$ condition, because
$\tau_{RC}=R_{ch}C_{\rm ox}\propto L^2$, making submicron devices
favorable for the measurement.
At high frequencies, careful rf de-embedding of contact resistance and
pad parasitics will be required to isolate the intrinsic $g_m(\omega)$
from extrinsic contributions.
Although the discussion is illustrated for a dual-gate geometry,
the mechanism is not specific to a particular gate architecture:
gate-all-around devices~\cite{PaperI,Mukesh2022_GAA_review} enhance
the electrostatic coupling and hence the magnitude of $p\chi_v$,
whereas dual-gate devices offer a more readily accessible
experimental platform.

Several other mechanisms can produce frequency-dependent $g_m$
or hysteresis and must be distinguished.
Interface states and fast charge exchange are known to degrade the
subthreshold swing, whereas border traps produce hysteresis and
threshold-voltage drifts~\cite{Illarionov2020_Ditreview,Knobloch2020_trapping}.
By contrast, the valley hysteresis predicted here arises from delayed
intervalley redistribution and does not introduce an additional
interface-trap capacitance; within the present model, the subthreshold
swing remains unchanged.
Charge trapping in dielectrics can produce $\tau$ values from
microseconds to seconds, but shows no layer-dependent transconductance sign reversal and
no dependence on $\Delta_{K\Gamma}$.
Applying compressive biaxial strain, which shifts $\Delta_{K\Gamma}$
predictably~\cite{PaperI,Johari2012}, would modify the valley
contribution in a controlled manner, providing a decisive test.
Deviations from the predicted Lorentzian line shape of
$\mathrm{Im}[g_m(\omega)]$ would directly signal distributed intervalley
relaxation channels beyond the minimal single-$\tau_{\rm iv}$ model,
opening a route to resolving the relaxation spectrum electrically.

\section{Conclusion}
\label{sec:conclusion}

In summary, we have shown that the valley population in multilayer
WSe$_2$ field-effect transistors acts as a dynamic internal state variable
with relaxation time $\tau_{\rm iv}$.
The resulting frequency-dependent transconductance
$g_m(\omega)=g_{m,0}+g_{m,v}^0/(1+i\omega\tau_{\rm iv})$,
two-stage transient response, and sweep-rate-proportional hysteresis
each provide distinct and electrically accessible probes of intervalley
relaxation dynamics.
The layer-dependent transconductance sign reversal and gate-voltage
profile of all three observables constitute robust fingerprints of
valley-thermodynamic origin, completing the dynamic extension of the
valley-thermodynamics framework for two-dimensional transistors~\cite{PaperII}.

\begin{acknowledgments}
This work was supported by JSPS KAKENHI (Grants No.~JP25K01609,
No.~JP22H05473, and No.~JP21H01019), JST CREST (Grant No.~JPMJCR19T1).
K.W. acknowledges financial support from the Sumitomo Foundation
(Grant No.~2401203).
\end{acknowledgments}

\section*{Data Availability}
The data used and analyzed during the current study are available from the corresponding author upon reasonable request.
\bibliography{references}

\end{document}